\documentclass[12pt]{article}
\usepackage{amsfonts,amssymb}
\usepackage{mathrsfs}

\def \d{\mathrm{d}}

\newcommand{\R}{{\mathbb{R}}}
\newcommand{\Z}{{\mathbb{Z}}}

\newcommand{\C}{{\mathbb{C}}}

\newcommand{\CP}{{\mathbb{C}}{{P}}}
\newcommand{\RP}{{\mathbb{R}{{P}}}}
\newcommand{\beq}{\begin{equation}}
\newcommand{\eeq}{\end{equation}}
\newcommand{\bea}{\begin{eqnarray}}
\newcommand{\eea}{\end{eqnarray}}
\newcommand{\ra}{\rightarrow}
\newcommand{\rat}{{\sf Rat}}
\newcommand{\h}{{\cal H}}

\newcommand{\hra}{\hookrightarrow}

\newcommand{\cd}{\partial}
\newcommand{\less}{\backslash}

\newcommand{\g}{{\mathfrak{g}}}
\renewcommand{\k}{{\mathfrak{k}}}
\newcommand{\p}{{\mathfrak{p}}}
\newcommand{\ignore}[1]{}
\newcommand{\ip}[1]{\langle #1 \rangle}

\newcommand{\tr}{{\rm tr}\, }

\newcommand{\vol}{{\rm vol}}
\newcommand{\diag}{{\rm diag}}

\begin{document}

\title{The volume of the space of holomorphic maps\\
from $S^2$ to $\CP^k$}
\author{
J.M. Speight\thanks{E-mail: {\tt speight@maths.leeds.ac.uk}}\\
School of Mathematics, University of Leeds\\
Leeds LS2 9JT, England
}

\date{}
\maketitle

\begin{abstract}
Let $\Sigma$ be a compact Riemann surface and $\h_{d,k}(\Sigma)$ denote the
space of degree $d\geq 1$ holomorphic maps $\Sigma\ra \CP^k$.
In theoretical physics this arises as the moduli space of charge $d$ lumps (or
instantons) in the $\CP^k$ model on $\Sigma$. There is a natural 
Riemannian metric
on this moduli space, called the $L^2$ metric,
 whose geometry is conjectured to control the
low energy dynamics of $\CP^k$ lumps. In this paper 
an explicit formula for the $L^2$ metric on
of $\h_{d,k}(\Sigma)$ in the special case $d=1$ and $\Sigma=S^2$
 is computed. Essential use is made of the k\"ahler property of the $L^2$ 
metric, and its 
invariance under a natural action of $G=U(k+1)\times U(2)$. It is shown
that {\em all} $G$-invariant k\"ahler metrics on $\h_{1,k}(S^2)$ 
have finite volume for $k\geq 2$. The volume of $\h_{1,k}(S^2)$ 
with respect to the $L^2$ metric is computed explicitly and is shown to
agree with a general formula for $\h_{d,k}(\Sigma)$
recently conjectured by Baptista. The area of a family of twice punctured 
spheres in $\h_{d,k}(\Sigma)$ is computed exactly, and a formal argument
is presented in support of Baptista's formula for $\h_{d,k}(S^2)$
for all $d$, $k$, and $\h_{2,1}(T^2)$.
%\vspace*{0.3cm}\newline
%PACS: \newline
%Keywords: 
\end{abstract}

\maketitle

\section{Introduction}
\label{intro}

Let $\Sigma$ be a compact Riemann surface of genus $g$ and $Y=\CP^k$ equipped
with the Fubini-Study metric of constant holomorphic sectional curvature
$c_2>0$. Maps $\phi:\Sigma\ra Y$ are classifed topologically by an 
integer degree $d=\int_\Sigma\phi^*\omega_Y$, where $\omega_Y$ is a suitably
normalized k\"ahler form on $Y$. It is well known that among all degree
$d\geq 1$ maps, the Dirichlet energy
$$
E(\phi)=\frac12\int_\Sigma\|\d\phi\|^2
$$
is minimized when $\phi$ is holomorphic. Let $\h_{d,k}$ denote the
set of degree $d$ holomorphic maps $\Sigma\ra Y$. In physics language, this
is the moduli space of charge $d$ $\CP^k$ lumps (or instantons) on $\Sigma$. 
If we equip
$\Sigma$ with a Riemannian metric (note that $E$ requires only a
conformal structure on $\Sigma$), then $\h_{d,k}$ inherits a
natural Riemannian metric $\gamma_{L^2}$ defined so that for any curve
$\phi(t)$ in $\h_{d,k}$,
$$
\gamma_{L^2}(\dot\phi,\dot\phi)=\int_\Sigma\|\dot\phi\|^2.
$$
This is usually called the $L^2$ metric, and may be intertpreted as (twice) 
the kinetic
energy of the time varying field $\phi(t,x)$. Following Manton's
approach to soliton dynamics \cite{mansut}, 
it is thought that geodesic motion in
$\h_{d,k}$ approximates classical low-energy $d$-lump dynamics on $\Sigma$.
The quantum energy spectrum of a $d$ lump system can be related to the
spectrum of an appropriate Laplace-Beltrami operator on $\h_{d,k}$, and
the equation of state of a classical gas of lumps can be deduced if one
knows the volume growth of $\h_{d,k}$ as a function of $d$. So given exact
information about the metric $\gamma_{L^2}$ there is a well-developed 
programme for 
extracting information about the classical, quantum and statistical
mechanics of lumps moving on $\Sigma$.

In this regard, Baptista \cite{bap2}
has recently made a very interesting conjecture
for the volume of $\h_{d,k}(\Sigma)$, motivated by a certain singular limit of 
a related abelian Yang-Mills-Higgs theory, namely, provided $d>2g-1$,
$$
{\rm Vol}(\h_{d,k}(\Sigma))=\frac{(k+1)^g}{N_{d,k,g}!}\left(\frac{4\pi}{c_2}
{\rm Vol}(\Sigma)\right)^{N_{d,k,g}},\qquad N_{d,k,g}=(k+1)(d+1-g)+g-1.
$$
In this paper we will make a detailed study of $\gamma_{L^2}$ in the case
$d=1$ and $\Sigma=S^2$ (with the round metric), confirming Baptista's
conjecture for all $k\geq 2$ (the case $k=1$ was already known by earlier
work of Baptista himself \cite{bap1}).  What makes
this case ($d=1$, $\Sigma=S^2$) tractable is that there is a cohomogeneity
1 isometric group action, built from the isometry groups of $S^2$ and $\CP^k$,
so we may decompose $\h_{1,k}$ into a one parameter family of homogeneous
orbits. The k\"ahler property then almost completely determines $\gamma_{L^2}$:
we will see that an arbitrary invariant k\"ahler metric on $\h_{1,k}$
is determined by a single function of one variable and a single constant.
It is not hard to show that all metrics with this structure have finite volume
for $k\geq 2$, and to find a formula for this volume.
This is a nontrivial and rather surprising result, given that
$\h_{1,k}$ is noncompact, and that $\h_{1,1}$ is known to admit
invariant k\"ahler metrics of infinite volume (one example is the 
Stenzel metric on $TS^3$, also known as the
``deformed conifold'' \cite{candel,stenzel,speconifold}). 

It turns out that $(\h_{1,k}(S^2),\gamma_{L^2})$ is geodesically incomplete.
This can be seen immediately from our explicit formula for the metric,
but actually follows in considerably more generality from previous work on 
the $\CP^1$ model. We note that there
is a totally geodesic inclusion
$
\iota:\h_{d,1}(\Sigma)\hra \h_{d,k}(\Sigma)
$
induced by the inclusion $\CP^1\hra\CP^{k}$, $[z_0,z_1]\mapsto
[z_0,z_1,0,\ldots,0]$. So it follows from the results of \cite{sadspe}
that whenever $\h_{d,1}(\Sigma)$ is nonempty,
$\h_{d,k}(\Sigma)$ is geodesically incomplete with respect to $\gamma_{L^2}$.
In particular, $\h_{d,k}(S^2)$ is incomplete for all $d,k$. So geodesic motion
in $\h_{d,k}(S^2)$ can hit the boundary at infinity in finite time, 
corresponding to
one or more lumps collapsing to zero width. It is likely, however, that the 
boundary at infinity has high codimension, so that generic geodesics
never hit it. (We shall see it has codimension $2k$ in the case
$\h_{1,k}(S^2)$.) If so, this means that Manton's
discussion \cite{manstatmech}
of the statistical mechanics of geodesic flow on moduli
space at large degree still makes sense (the set of bad initial data has
measure $0$), and we can hope to derive an equation of state for a gas
of $d$ lumps moving on a sphere of total area $A$. If Baptista's 
conjecture is correct, this equation of state turns out to be the
ideal gas equation
$$
PA=(k+1)dT,
$$
where $P$ is pressure and $d$ is interpreted as the number of lumps in the
gas.

The rest of this paper is structured as follows. In section \ref{metric}
we analyze the structure of invariant k\"ahler metrics on $\h_{1,k}(S^2)$
and find an explicit formula for $\gamma_{L^2}$. In section \ref{volume}
we show that the volume of $\h_{1,k}(S^2)$ with respect to an aribitrary
invariant k\"ahler metric is finite, and compute the volume with respect to
$\gamma_{L^2}$, confirming Baptista's conjecture
in these cases. Finally, section
\ref{dodgy} presents more indirect evidence in favour of Baptista's
conjecture for $\h_{d,k}(S^2)$, $d\geq 2$, $k\geq 1$ and $\h_{2,1}(T^2)$.

\section{The metric}
\label{metric}

Throughout the next two sections, $\Sigma=S^2$ and $d=1$.
It is convenient to identify the domain $\Sigma$ with $\CP^1$ given the
Fubini-Study metric of holomorphic sectional curvature $c_1$ 
(equivalently, $S^2$ given
the round metric of radius $1/\sqrt{c_1}$). A degree 1 holomorphic map $\CP^1
\ra \CP^k$ is one which lifts to a rank 2 linear map $\C^2\ra\C^{k+1}$, so
we have an identifcation of $\h_{1,k}$ with the set of 
projective equivalence classes of rank 2 $(k+1)\times 2$ complex matrices,
explicitly,
\beq
\phi([z_0,z_1])=[a_0z_0+b_0z_1,\ldots,a_kz_0+b_kz_1]\leftrightarrow
[M_\phi]=\left[\left(\begin{array}{cc}a_0&b_0\\ \vdots &\vdots \\a_k&b_k
\end{array}\right)\right].
\eeq
We may further identify $[M_\phi]$ with $[a_0,\ldots,a_k,b_0,\ldots,b_k]\in
\CP^{2k+1}$ to obtain an open inclusion $\h_{1,k}\hra \CP^{2k+1}$ whose image
is the complement of a complex codimension $k$ variety biholomorphic
to $\CP^1\times\CP^k$ (corresponding to the rank 1 matrices). The 
biholomorphism is
\beq
([x_0,x_1],[y_0,\ldots,y_k])\mapsto
[x_0y_0,\ldots,x_0y_k,x_1y_0,\ldots,x_1y_k].
\eeq
The inclusion into $\CP^{2k+1}$
equips $\h_{1,k}$ with a complex structure. By a straightforward
extension of the argument in \cite{speL2}, the $L^2$ metric on $\h_{1,k}$
is k\"ahler with respect to this complex structure. 

There is a natural
left action of $G=U(k+1)\times U(2)$ on $\h_{1,k}$ given 
by
\beq
(U_1,U_2):[M_\phi]\mapsto[U_1M_\phi U_2^{-1}].
\eeq
This maps $\phi$ to $i_2\circ\phi\circ i_1$ where $i_1,i_2$ are
isometries of $\CP^1,\CP^k$ respectively. 
Hence the $L^2$ metric is invariant under this
$G$ action.
The action has cohomogeneity 1, meaning that generic orbits are
codimension 1 submanifolds of $\h_{1,k}$. Each orbit contains
a unique map of the form
\beq
\phi_\mu([z_0,z_1])=[\mu z_0,z_1,0,\ldots,0],\qquad \mu\geq 1
\eeq
so the action decomposes $\h_{1,k}$ into a one parameter family of
orbits parametrized by $\mu\in[1,\infty)$, each orbit diffeomorphic to $G/K$,
where $K$ is the isotropy group of $\phi_\mu$. For $\mu>1$ (the
single exceptional orbit $\mu=1$, which has codimension
$3$, will not concern us), one finds that
$K$ is isomorphic to $T^3\times U(k-1)$, one isomorphism being
\beq\label{Kdef}
(e^{i\xi},e^{i\alpha},e^{i\beta},U)\mapsto
\left(
 \left(
  \begin{array}{ccc}
    e^{i\alpha}&0&0\\
    0&e^{i\beta}&0\\
    0&0& U
  \end{array}
 \right),
 \left(
  \begin{array}{cc}
    e^{i(\alpha+\xi)}&0\\0&e^{i(\beta+\xi)}
  \end{array}
 \right)
\right).
\eeq

From now on, let $\gamma$ be any $G$-invariant hermitian metric on $\h_{1,k}$.
Let $\g,\k$ be the Lie algebras of $G,K$ respectively, and denote
by $\ip{,}$ the natural $Ad(G)$ invariant inner product on $\g$, namely
\beq\label{ip}
\ip{(A,B),(A',B')}=-\frac12(\tr AA'+\tr BB').
\eeq
We may identify the tangent space to the orbit through
$\phi_\mu$ with any $Ad(K)$ invariant subspace $\p$ complementary to
$\k$ in $\g$. We choose $\p=\k^\perp$, the orthogonal complement to
$\k$ with respect to $\ip{,}$. Any other tangent space $T_{gK}(G/K)$
may be identified with $\p$ by left translation by $g^{-1}$, but this
identification is only well-defined modulo the adjoint action of $K$ on $\p$,
since the element $u\in\p$ with which $X\in T_{gK}(G/K)$ is identified
moves on an $Ad(K)$ orbit as $g$ takes all values in $gK$. It follows that
the metric $\gamma$ on $\h_{1,k}$ is uniquely 
determined by the one-parameter family of symmetric bilinear forms
\beq
\gamma_\mu: V_\mu\times V_\mu\ra\R
\eeq
where $V_\mu=T_{\phi_\mu}\h_{1,k}=\ip{\cd/\cd\mu}\oplus\p$, and each
of these bilinear forms must be invariant under the adjoint action of
$K$ on $\p$. Since $\gamma$ is assumed to be hermitian,
$\gamma_\mu$ must also be invariant under
the action of the almost complex structure $J$ on $V_\mu$. 

This leads us to decompose $\p$
into subspaces
\beq
\p=\p_0\oplus\p_\mu\oplus\tilde\p_\mu\oplus\hat\p\oplus\check\p
\eeq
\ignore{of real dimension $1,2,2,2(k-1),2(k-1)$ respectively,} defined by
\bea\label{p0}
\p_0&=&\left\{\lambda(\diag(i,-i,0,\ldots,0),\diag(-i,i))\: :\: \lambda\in\R\right\}\equiv
\R\\
\label{pmu}
\p_\mu&=&
\left\{
 \left(
  \left(
   \begin{array}{cccc}
    0&x&0&\cdots\\-\bar{x}&0&0&\cdots\\ 0&0&&\\\vdots&\vdots&&
   \end{array}
  \right),
  \left(
   \begin{array}{cc}
    0&\mu x\\-\mu\bar{x}&0
   \end{array}
  \right)
 \right)
 \: :\: x\in\C
\right\}\equiv\C\\
\label{pmutilde}
\tilde{\p}_\mu&=&
\left\{
 \left(
  \left(
   \begin{array}{cccc}
    0&-\mu\bar{y}&0&\cdots\\ \mu y&0&0&\cdots\\0&0&&\\\vdots&\vdots&&
   \end{array}
  \right),
  \left(
   \begin{array}{cc}
    0&-\bar{y}\\y&0
   \end{array}
  \right)
 \right)
 \: :\: y\in\C
\right\}\equiv\C\\
\label{phat}
\hat\p&=&\left\{\left(\left(\begin{array}{cc}
\begin{array}{cc}0&0\\0&0\end{array}&\begin{array}{c}-u^\dagger\\0\end{array}\\
\begin{array}{cc}u&0\end{array}&0\end{array}\right),0\right)\: :\:
u\in\C^{k-1}\right\}\equiv\C^{k-1}\\
\label{pcheck}
\check\p&=&\left\{\left(\left(\begin{array}{cc}
\begin{array}{cc}0&0\\0&0\end{array}&\begin{array}{c}0\\-v^\dagger\end{array}\\
\begin{array}{cc}0&v\end{array}&0\end{array}\right),0\right)\: :\:
v\in\C^{k-1}\right\}\equiv\C^{k-1}
\eea
With respect to this decomposition, the action of $J$ on $(\lambda,x,y,u,v)\in\p$
is
\beq
J:(\lambda,x,y,u,v)\mapsto 4\mu\lambda\frac{\cd\,}{\cd\mu}+(0,ix,iy,iu,iv),
\eeq
and the adjoint action of $K\equiv T^3\times U(k-1)$ is
\beq
(e^{i\xi},e^{i\alpha},e^{i\beta},U):(\lambda,x,y,u,v)\mapsto
(\lambda,e^{i(\alpha-\beta)}x,e^{-i(\alpha-\beta)}y,e^{-i\alpha}Uu,e^{-i\beta}Uv).
\eeq
Hence, any $G$-invariant hermitian metric $\gamma$ on $\h_{1,k}$ has
\beq\label{herm}
\gamma_\mu=A_0(\mu)(d\mu^2+8\mu^2\ip{,}_{\p_0})+
A_1(\mu)\ip{,}_{\p_\mu}+
A_2(\mu)\ip{,}_{\tilde{\p}_\mu}+
A_3(\mu)\ip{,}_{\hat{\p}}+
A_4(\mu)\ip{,}_{\check{\p}}
\eeq
where $A_0,\ldots,A_4$ are smooth positive functions of $\mu$.

Now let us assume further that the metric $\gamma$ is k\"ahler. By previous work
of Dancer and Wang \cite{danwan}, this implies the following constraints on 
the k\"ahler
form $\omega(\cdot,\cdot)=\gamma(J\cdot,\cdot)$:
\bea
\label{k1}
\omega([X,Y]_\p,Z)+\omega([Y,Z]_\p,X)+\omega([Z,X]_\p,Y)&=&0\\
\label{k2}
\frac{\cd\,}{\cd \mu}\omega(X,Y)-\omega(\frac{\cd X}{\cd\mu},Y)-
\omega(X,\frac{\cd Y}{\cd\mu})+\omega(\frac{\cd\,}{\cd\mu},[X,Y]_\p)&=&0
\eea
where $X,Y,Z$ are any (possibly $\mu$ dependent) elements of $\p$.
Constraint (\ref{k1}) in the case $X=(0,1,0,0,0)\in\p_\mu$, $Y=(0,0,1,0,0)\in\tilde\p_\mu$
and $Z=(1,0,0,0,0)\in\p_0$ implies that 
\beq
A_1(\mu)=A_2(\mu).
\eeq
 In the case
$X=(0,1,0,0,0)\in\p_\mu$,
$Y=(0,0,0,(1,0,\ldots,0),0)\in\hat\p$ and
$Z=(0,0,0,0,(i,0,\ldots,0))\in\check\p$, constraint (\ref{k1}) implies
\beq
A_4(\mu)-\frac{\mu^2+1}{\mu^2-1}A_1(\mu)-A_5(\mu)=0.
\eeq
Turning to constraint (\ref{k2}), the choice $X=(0,1,0,0,0)\in\p_\mu$, 
$Y=JX=(0,i,0,0,0)\in\p_\mu$, yields
\beq
A_0=\frac{1}{4\mu}\frac{\d\:}{d\mu}\left(\frac{\mu^2+1}{\mu^2-1}A_1\right),
\eeq
while $X=(0,0,0,(1,0\ldots,0),0)\in\hat\p$, $Y=JX=(0,0,0,(i,0\ldots,0),0)\in\hat\p$,
yields
\beq
\frac{dA_4}{d\mu}-2\mu A_0=0
\eeq
and
 $X=(0,0,0,0,(1,0\ldots,0))\in\check\p$, $Y=JX=(0,0,0,0,(i,0\ldots,0))\in\check\p$,
yields
\beq
\frac{dA_5}{d\mu}+2\mu A_0=0.
\eeq

Assembling these constraints, we see that any $G$-invariant k\"ahler metric
on $\h_{1,k}$ takes the form prescribed in (\ref{herm}) with the coefficient functions 
$A_0,\ldots,A_4$
uniquely determined by a smooth positive function $A(\mu)$, $\mu>1$
and a positive constant $B$, so that
\beq\label{gam}
A_0=\frac{1}{4\mu}\frac{dA}{d\mu},\quad
A_1=A_2=\frac{\mu^2-1}{\mu^2+1}A,\quad
A_3=B+\frac12A,\quad
A_4=B-\frac12A
\eeq
Given that $\gamma$ is positive definite, $A$ must be strictly increasing and
bounded above by $2B$. Hence $A$ has a limit $A(\infty)\leq 2B$. Also, since 
$\gamma$ extends to the exceptional orbit $\mu=1$, one finds that $A(1)=\lim_{\mu\ra 1}
A(\mu)=0$.

The analysis above applies to all $G$-invariant k\"ahler metrics on $\h_{1,k}$, of
which the $L^2$ metric is an example. To construct $\gamma_{L^2}$ completely,
it remains to compute $A(\mu)$ and $B$.
An economical way to do this is to compute the squared lengths
of $X=(0,0,0,(1,0,\ldots,0),0\in\hat\p$ and $Y=(0,0,0,0(1,0,\ldots,0))\in\check\p$.
For the $L^2$ metric, one finds that
\beq\label{L2}
A_{L^2}(\mu)=\frac{16\pi}{c_1c_2}\, \frac{\mu^4-4\mu^2\log\mu-1}{(\mu^2-1)^2},\qquad
B_{L^2}=\frac{8\pi}{c_1c_2},
\eeq
where $c_1,c_2$ are the holomorphic sectional curvatures of the domain, $\CP^1$, and
target, $\CP^k$, respctively \cite{kobnom2}. 
An elementary estimate shows that the length of the curve $\phi_\mu$,
$\mu\geq 1$ is finite, whence it follows that this metric has finite diameter
and is incomplete.

Recall that $\h_{1,k}$ sits naturally as
an open subset of $\CP^{2k+1}$, whence it inherits an alternative k\"ahler metric
$\gamma_{FS}$, induced by the Fubini-Study metric of holomorphic sectional curvature
$c$, say, on $\CP^{2k+1}$. 
This metric is invariant under the natural action of $G'=U(2k+2)$, which
contains $G$ as a subgroup,
$G\hra G'$, $(U_1,U_2)\mapsto U_1\otimes U_2^{-1}$. Hence
this metric is also $G$-invariant, so has the structure
prescribed by equations (\ref{herm}), (\ref{gam}). In 
this case
\beq
A_{FS}(\mu)=\frac{4}{c}\left(\frac{\mu^2-1}{\mu^2+1}\right),\qquad 
B_{FS}=\frac{2}{c}.
\eeq
Note that for both $\gamma_{L^2}$ and $\gamma_{FS}$, $A(\mu)$ is an increasing function
from $A(1)=0$ to $A(\infty)\leq 2B$, as required for positivity and
regularity. In fact, $A(\infty)=2B$ in both cases, so as $\mu\ra\infty$ the
subspace $\ip{\cd/\cd\mu}\oplus\p_0\oplus\check\p$ of $T_{\phi_\mu}\h_{1,k}$ collapses,
and the boundary of $\h_{1,k}$ at infinity has (real) codimension
$2k$. This is consistent with our earlier observation that the
rank 1 matrices form a complex codimension $k$ submanifold  
of $\CP^{2k+1}$. 

\section{The volume of $\h_{1,k}$}
\label{volume}

In this section we will show that {\em every}
$G$-invariant k\"ahler metric on $\h_{1,k}$ ($k\geq 2$) has finite total
volume,
a result in stark contrast to the previously considered
case $k=1$, where invariant k\"ahler metrics of infinite volume certainly exist
\cite{speconifold,bap1}.
We will also find a formula for this volume in terms of $B$ and $k$
under the extra assumption that $A(\infty)=2B$, as holds for the $L^2$ metric. 
This formula confirms Baptista's conjecture in the cases under consideration.

We start by computing the volume form of any $G$-invariant hermitian metric
on $\h_{1,k}$. Denote by $\vol_{G/K}$ the volume form on
$G/K$ induced by the $Ad(G)$ invariant metric $\ip{,}$, and $\vol_0$ the 
volume form on $(1,\infty)\times G/K$ induced by the product metric $\gamma_0=d\mu^2+
\ip{,}_\p$, so $\vol_0=d\mu\wedge\vol_{G/K}$. Then, the volume form induced by $\gamma$
is, for some smooth positive function $F(\mu)$,
\beq\label{1}
\vol=F(\mu)\vol_0,
\eeq
and we seek to deduce $F(\mu)$. To do this, we construct orthonormal bases for
$\gamma$ and $\gamma_0$. Let $e_{ij}$ denote a square matrix (the size will be
either $2\times 2$ or $(k+1)\times(k+1)$, which being clear from context) with $(i,j)$ entry
$1$ and all others $0$. Then an orthonormal basis for $\gamma_0$ is given by
\bea
&&Y_1=\frac{\cd\:}{\cd\mu},\quad Y_2=\frac{1}{\sqrt{2}}(ie_{11}-ie_{22},-ie_{11}+ie_{22}),
\nonumber \\
&&Y_3=(e_{12}-e_{21},0),\quad Y_4=(ie_{12}+ie_{21},0),\quad  
Y_5=(0,-e_{12}+e_{21}),\quad Y_6=(0,ie_{12}+ie_{21}), \nonumber \\
&&\hat{Y}_{2j-1}=(-e_{1,j+2}+e_{j+2,1},0),\quad
\hat{Y}_{2j}=(ie_{1,j+2}+ie_{j+2,1},0),\quad j=1,\ldots,k-1 \nonumber \\
\label{Ydef}
&&\check{Y}_{2j-1}=(-e_{2,j+2}+e_{j+2,2},0),\quad
\check{Y}_{2j}=(ie_{2,j+2}+ie_{j+2,2},0),\quad j=1,\ldots,k-1
\eea
and an orthonormal basis for $\gamma$ is given by
\bea
&&X_1=\frac{1}{\sqrt{A_0}}Y_1,\quad X_2=\frac{1}{\mu\sqrt{8A_0}}Y_2,\nonumber \\
&&X_3=\frac{Y_3-\mu Y_5}{\sqrt{(1+\mu^2)A_1}},\quad
X_4=\frac{Y_4+\mu Y_6}{\sqrt{(1+\mu^2)A_1}},\quad
X_5=\frac{-\mu Y_3+Y_5}{\sqrt{(1+\mu^2)A_2}},\quad
X_6=\frac{\mu Y_4+Y_6}{\sqrt{(1+\mu^2)A_2}},\nonumber \\
\label{Xdef}
&&\hat{X}_j=\frac{1}{\sqrt{A_3}}\hat{Y}_j,\quad
\check{Y}_j=\frac{1}{\sqrt{A_4}}\check{Y}_j,\quad j=1,\ldots,2k-2.
\eea
Evaluating both sides of (\ref{1}) on the orthornormal basis for $\gamma$ gives
\bea
1&=&F(\mu)\vol_0(X_1,\ldots,X_6,\hat{X}_1,\ldots,\hat{X}_{2k-2},\check{X}_1,\ldots,
\check{X}_{2k-2})\nonumber \\
&=&\frac{F(\mu)\vol_0(Y_1,Y_2,Y_3-\mu Y_5,Y_4+\mu Y_6,-\mu Y_3+Y_5,\mu Y_4+Y_6,\hat{Y}_1,\ldots,
\check{Y}_1,\ldots)}{\sqrt{8}\mu(1+\mu^2)^2A_0A_1A_2(A_3A_4)^{k-1}}
\nonumber \\
&=&\frac{(1-\mu^2)^2F(\mu)}{\sqrt{8}\mu(1+\mu^2)^2A_0A_1A_2(A_3A_4)^{k-1}}.
\eea
Hence, the volume form of a general $G$-invariant hermitian metric on $\h_{1,k}$ is
\beq
\vol=\sqrt{8}\mu\left(\frac{\mu^2+1}{\mu^2-1}\right)^2A_0A_1A_2(A_3A_4)^{k-1} d\mu\wedge
\vol_{G/K}.
\eeq

Assume now that the metric is k\"ahler. Then, owing to (\ref{gam}),
\beq
\vol=\frac{1}{\sqrt{2}}A^2(B^2-\frac{A^2}{4})^{k-1}\frac{dA}{d\mu}\: d\mu\wedge
\vol_{G/K},
\eeq
whence we find that the total volume of $\h_{1,k}$ is
\bea
{\rm Vol}(\h_{1,k})&=&\frac{1}{\sqrt{2}}\int_1^\infty A^2(B^2-\frac{A^2}{4})^{k-1}\frac{dA}{d\mu}\: d\mu\int_{G/K}\vol_{G/K}\nonumber \\
&=&4\sqrt{2}B^{2k+1}{\rm Vol}(G/K)\int_0^{A(\infty)/2B}t^2(1-t^2)^{k-1}\, dt,
\eea
which is finite. Hence, every $G$-invariant k\"ahler metric on $\h_{1,k}$
has finite volume (for $k\geq 2$).

 Let us assume further that $A(\infty)=2B$, as holds for both the
$L^2$ metric and the Fubini-Study metric. Then
\beq
{\rm Vol}(\h_{1,k})=B^{2k+1}\alpha_k
\eeq
where
\beq
\alpha_k=4\sqrt{2}{\rm Vol}(G/K)\int_0^{1}t^2(1-t^2)^{k-1}\, dt
\eeq
depends only on $k$. It is not hard to compute the above integral exactly.
 However,
computing ${\rm Vol}(G/K)$ (which also depends on $k$, of course) is not so easy, so
we deduce $\alpha_k$ indirectly, as follows. Let $\gamma_{FS}$ be the Fubini-Study
metric on $\h_{1,k}$ of holomorphic sectional curvature $c=1$, and hence
with
$B_{FS}=2$. Since the complement of $\h_{1,k}$ in $\CP^{2k+1}$ has
measure $0$ (it is a codimension $k$ submanifold), the volume of $\h_{1,k}$ 
with 
respect to $\gamma_{FS}$ coincides with the volume of $\CP^{2k+1}$. 
But this volume is well known to be $(4\pi)^{2k+1}/(2k+1)!$ (this follows 
\cite{manstatmech} from the fact
that the integral of the k\"ahler form over any $\CP^1$ submanifold
generating $H_2(\CP^{2k+1},\Z)$ is $4\pi$).
 Hence, for this particular Fubini-Study metric
\beq
{\rm Vol}(\h_{1,k})=2^{2k+1}\alpha_k=\frac{(4\pi)^{2k+1}}{(2k+1)!}
\eeq
whence we deduce that 
\beq
\alpha_k=\frac{(2\pi)^{2k+1}}{(2k+1)!}.
\eeq
Hence, the volume of $\h_{1,k}$ with respect to any $G$-invariant k\"ahler 
metric
with $A(\infty)=2B$ is
\beq
{\rm Vol}(\h_{1,k})=\frac{(2B\pi)^{2k+1}}{(2k+1)!}.
\eeq
For the $L^2$ metric, one sees from (\ref{L2}) that for any $k\geq 2$.
\beq
{\rm Vol}_{L^2}(\h_{1,k})=\frac{1}{(2k+1)!}\left(\frac{4\pi}{c_1}\, \frac{4\pi}{c_2}\right)^{2k+1}.
\eeq
This confirms Baptista's conjecture for these
moduli spaces. We note that the same formula is known to
hold in the case $k=1$ (see \cite{bap1}, which uses the convention $c_1=c_2=4$). 

In fact, we can compute exactly the volume of {\em any} $G$-invariant k\"ahler 
metric
on $\h_{1,k}$, even if $A(\infty)<2B$. Since we know $\alpha_k$, and 
\beq
\int_0^1 t^2(1-t^2)^{k-1}dt=\frac{(k-1)!2^{k-1}}{(2k+1)!!}
\eeq
we deduce that
\beq
{\rm Vol}(G/K)=\frac{2^k\pi^{2k+1}}{\sqrt{2}(k-1)!k!},
\eeq
and hence
\beq
{\rm Vol}(\h_{1,k})=\frac{2^{k+2}(B\pi)^{2k+1}}{(k-1)!k!}\int_0^{A(\infty)/2B}t^2(1-t^2)^{k-1}dt.
\eeq
The last integral can easily be computed explicitly using the binomial theorem,
but the final answer is not very instructive. It is interesting to
note that the total volume depends only on the asymptotic behaviour of
the metric close to the boundary at infinity.

\section{Higher degree and genus}
\label{dodgy}

For higher degree $d$, or domain $\Sigma$ of higher genus $g$, the $L^2$
geometry of $\h_{d,k}(\Sigma)$ is much less accessible. Nonetheless, we can
make one exact calculation which, while not directly confirming 
Baptista's conjecture, seems to support it. 

Let $W$ be a meromorphic function of degree $d$
on a compact Riemann surface $\Sigma$. Associated to $W$ is a cylindrical
submanifold $C_W$ of $\h_{d,k}(\Sigma)$, consisting of the holomorphic
maps $\phi_\mu:\Sigma\ra\CP^k$ defined locally by 
$\phi_\mu(z)=[\mu W(z),1,0,\ldots,0]$, $\mu\in\C^\times=\C\less\{0\}$.
Physically, this is the orbit of a fixed $d$-lump under dilation
(changing $|\mu|$) and isorotation (changing ${\rm arg}(\mu)$).
The induced $L^2$ metric on $C_W$ is
\beq
\gamma|_{C_W}=F(\mu)\d\mu\d\bar\mu,\qquad
F(\mu)=\frac{4}{c_2}\int_\Sigma\frac{|W|^2}{(1+|\mu|^2|W|^2)^2},
\eeq
where the measure on $\Sigma$ is the one defined by its Riemannian metric.
Hence, the total volume of $C_W$ is the integral
\beq
{\rm Vol}(C_W)=\int_{\C^\times}F=\frac{4}{c_2}\int_{\C^\times}\left(
\int_\Sigma\frac{|W|^2}{(1+|\mu|^2|W|^2)^2}\right)
\eeq
if this exists (i.e., is finite). By Fubini's theorem \cite{fub}, the integral
exists if and only if
\beq\label{10}
\int_{\Sigma}\left(
\int_{\C^\times}\frac{|W|^2}{(1+|\mu|^2|W|^2)^2}\right)
\eeq
exists, in which case they are equal. But (\ref{10}) is trivial:
\bea
\int_{\Sigma}\left(
\int_{\C^\times}\frac{|W|^2}{(1+|\mu|^2|W|^2)^2}\right)
&=&2\pi\int_{\Sigma}\left(\int_0^\infty d|\mu|\frac{|\mu||W|^2}{(1+|\mu|^2|W|^2)^2}\right)
=2\pi\int_\Sigma\frac12.
\eea
Hence, the cylinder $\C^\times_W$ has total volume
\beq\label{em}
{\rm Vol}(C_W)=\frac{4\pi}{c_2}{\rm Vol}(\Sigma).
\eeq
Note that this is independent of the meromorphic function $W$.
The above calculation of ${\rm Vol}(C_W)$ generalizes the result in
\cite{macspe}, which considered
the case $k=1$, $\Sigma=S^2$ and $W=z^d$ 
(where $z$ is a stereographic coordinate on $S^2$). In that case,
$C_W$ is a totally geodesic submanifold of $\h_{d,k}$, but in general
there is no reason why $\C_W^\times$ should be totally geodesic.

We emphasize that all the results of sections \ref{metric} and
\ref{volume}, and the present section up to this point, are
mathematically rigorous. The remaining paragraphs of this section
are suggestive, rather than rigorous.

The formula (\ref{em}) supports Baptista's conjecture as follows. Let
$\Sigma=\CP^1$ (with any metric). Then points in $\h_{d,k}$ 
may be identified with projective equivalence classes of
$(k+1)\times(d+1)$ complex matrices via
\beq
\phi([z_0,z_1])=[a_{00}z_0^d+a_{01}z_0^{d-1}z_1+\cdots+
a_{k,d}z_1^d,\ldots,
a_{k0}z_0^d+a_{01}z_0^{d-1}z_1+\cdots+
a_{k,d}z_1^d]
\leftrightarrow
\left[\left(a_{ij}\right)\right].
\eeq
This gives an open inclusion $\h_{d,k}\hra\CP^{dk+d+k}$ whose complement is
again an algebraic variety of high codimension. Suppose that the $L^2$ metric
on $\h_{d,k}$ extends smoothly to $\CP^{dk+d+k}$. Then the $L^2$ volume of
$\h_{d,k}$ coincides with the $L^2$ volume of $\CP^{dk+d+1}$,
\beq\label{11}
{\rm Vol}(\h_{d,k})=\int_{\CP^{dk+d+k}}\frac{\omega_{L^2}^{dk+d+k}}{(dk+d+k)!}
=\frac{1}{(dk+d+k)!}\left(\int_{X}\omega_{L^2}\right)^{dk+d+k}
\eeq
where $X$ is any 2-cycle generating $H_2(\CP^{dk+d+k},\Z)$
\cite{manstatmech}. One such 2-cycle is
the cylinder $C_W$ completed by adding the points $\mu=0$, $\mu=\infty$,
where $W$ is the meromorphic function $(z_0/z_1)^d$. This is a complex 
submanifold of $\CP^{dk+d+k}$, so 
\beq
\int_X\omega_{L^2}={\rm Vol}(X)=\frac{4\pi}{c_2}{\rm Vol}(\Sigma)
\eeq
by (\ref{em}). Hence, {\em if $\gamma_{L^2}$ extends smoothly to
$\CP^{dk+d+k}$} then, when $g=0$,
\beq
{\rm Vol}(\h_{d,k})=\frac{1}{(dk+d+k)!}\left(\frac{4\pi}{c_2}{\rm Vol}(\Sigma)\right)^{dk+d+k},
\eeq
which agrees with Baptista's conjecture. 
Unfortunately, it is probable that $\gamma_{L^2}$ {\em never} extends smoothly
to $\CP^{dk+d+1}$. Certainly it cannot when $d=1$, as the scalar curvature
of $\gamma_{L^2}$ is unbounded in this case. So the above argument is purely
formal. It may be, however, that
the metric extends sufficiently regularly for the crucial step (\ref{11})
above to make sense. 

Finally, let us consider the case of genus $g=1$, that is, $\Sigma=T^2$, with
$k=1$ (target $\CP^1$) and degree $d=2$. 
It is known \cite{speT2} that there is a four-fold covering map
\beq
\rat_1\times T^2\ra \h_{2,1}(T^2),\qquad
(R,z_0)\mapsto \phi(z)=[R(\wp(z-z_0)),1]
\eeq
where $\wp$ is the Weierstrass P function, and $\rat_1$ denotes the space of 
degree 1 rational maps (fractional linear transformations).
 The $L^2$ metric on $\h_{2,1}(T^2)$ lifts to a k\"ahler
product metric on $\rat_1\times T^2$, the $T^2$ factor being
$8\pi/c_2$ (the rest mass of a charge $2$ lump)
times the metric on $\Sigma$. So,
\beq
{\rm Vol}(\h_{2,1}(T^2))=\frac14\, \frac{8\pi}{c_2}{\rm Vol}(\Sigma)
{\rm Vol}(\rat_1).
\eeq
It is not known whether $\rat_1$ has finite volume in this geometry. However,
repeating the formal argument above, we can compactify $\rat_1$ to obtain
$\CP^3$, and compute its volume as $({\rm Vol}(C_W))^3/3!$, where
$W=\wp$. But, by (\ref{em}), 
${\rm Vol}(C_W)=4\pi c_2^{-1}{\rm Vol}(\Sigma)$. Hence, we are led to
expect that
\beq
{\rm Vol}(\h_{2,1}(T^2))=\frac{2}{4!}(4c_2^{-1}\pi{\rm Vol}(T^2))^4,
\eeq
which again coincides with Baptista's conjecture.

\subsection*{Acknowledgements} The author thanks Joao Baptista for
useful correspondence.

\end{document}